\documentclass[aps,prl,twocolumn,superscriptaddress,letterpaper,floatfix]{revtex4}
\usepackage{graphicx}
\usepackage{epsfig}
\usepackage{subfigure}
\usepackage{bm}

\newcommand{\greac}{\bar{p}p\rightarrow\bar{Y}Y}
\newcommand{\reac}{\mbox{$\bar{p}p\rightarrow\bar{\Lambda}\Lambda$}}
\newcommand{\lbar}{\bar{\Lambda}}
\newcommand{\lbarl}{\bar{\Lambda}\Lambda}
\newcommand{\pbarp}{\bar{p}p}
\newcommand{\sigv}{ \vec{\bm \sigma}}

\newcommand{\nh}{\hat{\bm n}}

\newcommand{\vp}{\vec{\bm p}}
\newcommand{\vpt}{\vec{\bm P}_{\bm T}}

\newcommand{\dnn}{D_{nn}}
\newcommand{\knn}{K_{nn}}
\newcommand{\pGeV}{\,\mbox{GeV/{\it c}}}
\newcommand{\sscm}{_{\mbox{\tiny cm}}}

\newcommand{\aCMU}{\affiliation{Carnegie Mellon University,  Pittsburgh, 
Pennsylvania 15213}}
\newcommand{\aJulich}{\affiliation{Institut f\"{u}r Kernphysik des 
Forschungszentrums J\"{u}lich, D-52425 J\"{u}lich, Germany}}
\newcommand{\aBochum}{\affiliation{Ruhr-Universit\"{a}t Bochum, 
D-44780 Bochum, Germany}}
\newcommand{\aBonn}{\affiliation{Universit\"{a}t Bonn, D-53115 Bonn, Germany}}
\newcommand{\aErlangen}{\affiliation{Universit\"{a}t Erlangen-N\"{u}rnberg, 
D-91058 Erlangen, Germany}}
\newcommand{\aFreiburg}{\affiliation{Universit\"{a}t Freiburg, 
D-79104 Freiburg, Germany}}
\newcommand{\aUIUC}{\affiliation{University of Illinois, 
Urbana, Illinois 61801}}
\newcommand{\aUNM}{\affiliation{University of New Mexico, Albuquerque, 
New Mexico 87131}}
\newcommand{\aUppsala}{\affiliation{Uppsala University, S-75121 Uppsala, Sweden}}

\begin{document}

\title{Measurement of Spin Transfer Observables in 
\mbox{$\bar{p}p\rightarrow\bar{\Lambda}\Lambda$} at 1.637$\pGeV$}

\collaboration{The PS185 Collaboration}\noaffiliation

\author{B.~Bassalleck}\aUNM
\author{A.~Berdoz}\aCMU
\author{C.~Bradtke}\aBochum
\author{R.~Br\"{o}ders}\aJulich
\author{B.~Bunker}\aUIUC 
\author{H.~Dennert}\aErlangen
\author{H.~Dutz}\aBonn
\author{S.~Eilerts}\aUNM
\author{W.~Eyrich}\aErlangen
\author{D.~Fields}\aUNM
\author{H.~Fischer}\aFreiburg
\author{G.~Franklin}\aCMU
\author{J.~Franz}\aFreiburg
\author{R.~Gehring}\aBochum
\author{R.~Geyer}\aJulich
\author{S.~Goertz}\aBochum
\author{J.~Harmsen}\aBochum
\author{J.~Hauffe}\aErlangen
\author{F.H.~Heinsius}\aFreiburg
\author{D.~Hertzog}\aUIUC
\author{T.~Johansson}\aUppsala
\author{T.~Jones}\aUIUC
\author{P.~Khaustov}\aCMU
\author{K.~Kilian}\aJulich
\author{P.~Kingsberry}\aUNM 
\author{E.~Kriegler}\aFreiburg
\author{J.~Lowe}\aUNM
\author{A.~Meier}\aBochum 
\author{A.~Metzger}\aErlangen
\author{C.A.~Meyer}\aCMU 
\author{W.~Meyer}\aBochum
\author{M.~Moosburger}\aErlangen
\author{W.~Oelert}\aJulich
\author{K.D.~Paschke}
	\altaffiliation[Now at ]{University of Massachusetts, Amherst 01003}
        \email{paschke@jlab.org}\aCMU 
\author{M.~Pl\"{u}ckthun}\aBonn
\author{S.~Pomp}\aUppsala
\author{B.~Quinn}\aCMU
\author{E.~Radtke}\aBochum
\author{G.~Reicherz}\aBochum
\author{K.~R\"{o}hrich}\altaffiliation[Now at ]{Creative Services, 
	F-01630 Saint Genis-Pouilly, France}\aJulich
\author{K.~Sachs}\aJulich
\author{H.~Schmitt}\aFreiburg
\author{B.~Schoch}\aBonn
\author{T.~Sefzick}\aJulich
\author{F.~Stinzing}\aErlangen
\author{R.~Stotzer}\aUNM
\author{R.~Tayloe}\altaffiliation[Now at ]{University of Indiana, 
	Bloomington, Indiana 47408}\aUIUC
\author{St.~Wirth}\aErlangen

\begin{abstract}
Spin transfer observables for the strangeness-production reaction $\reac$
have been measured by the PS185 collaboration using a transversely-polarized
frozen-spin target with an antiproton beam 
momentum of $1.637\pGeV$ at the Low Energy Antiproton 
Ring at CERN. 
This measurement investigates observables for which current models of the
reaction near threshold make significantly differing predictions.  
Those models are in good agreement with existing measurements performed
with unpolarized particles in the initial state.  Theoretical attention
has focused on the fact that these models produce conflicting
predictions for the spin-transfer observables $\dnn$ and $\knn$, which
are measurable only with polarized target or beam.
Results presented here for $\dnn$ and $\knn$ 
are found to be in disagreement with predictions from existing models.
These results also underscore the importance of singlet-state
production at backward angles, while current models predict 
complete or near-complete triplet-state dominance.
\end{abstract}

\maketitle

The measurement of near-threshold exclusive antihyperon-hyperon 
($\bar{Y}Y$) production in antiproton-proton reactions has 
proven to be a powerful tool in the study of the dynamics of the $\bar{q}q$
annihilation and production mechanisms.  
These reactions have been
extensively studied by the PS185 collaboration at the 
Low Energy Antiproton Ring (LEAR) facility at CERN, which
has produced the overwhelming majority of the 
existing data on exclusive $\greac$ near-threshold, including 
cross-sections and final-state polarization and 
spin-correlations~\cite{PS185M,PS185_96}.  
In particular, high precision
measurements have made of $\lbarl$ production over a kinematic
range from very-near threshold to about 200 MeV of excess energy in the 
center-of-mass system.  

The features of the $\reac$ reaction have been reproduced
by various models which describe the reaction dynamics either 
in a meson-exchange framework~\cite{Tabakin,LaFrance,Timmermans,Haidenbauer} 
or with a QCD-inspired effective theory~\cite{Furui,Khono,Alberg93}.
In the meson-exchange model (MEX), the transition occurs through the 
t-channel exchange
of a strange meson, with the $K(494)$ and $K^{*}(892)$ most often found to
provide the most significant contributions~\cite{Timmermans,Haidenbauer}.  
The QCD-inspired ``Quark-Gluon'' model (QG) is an effective theory 
describing an s-channel exchange with 
well-defined quantum numbers corresponding to specific QCD 
degrees-of-freedom, such as $^3S_1$ (single gluon exchange) 
or $^3P_0$ (multiple-gluon exchange with the quantum numbers of the 
vacuum)~\cite{Furui,Khono,Alberg93}. In both formulations, the initial-state 
and final-state interactions 
contribute significantly to the features of the reaction.  Although the
initial state interactions (ISI) are constrained by $\pbarp$ elastic
and inelastic scattering data, no similar data exists to constrain
the $\lbarl$ final-state interaction (FSI). 

Given the uncertainty in initial- and final-state interactions, as well as 
freedom in adjusting coupling strengths, calculations based on each of these two
approaches have successfully reproduced the previously measured observables,
although these calculations are quite different in 
predictions of the reaction dynamics~\cite{Timmermans,Haidenbauer,Alberg93}.  
One prominent disagreement between the models is the role
of the tensor interaction. MEX calculations lead to a dominant tensor force
resulting from a constructive interference of the $K$ and $K^*$
in the tensor channel. This tensor interaction couples only to 
triplet ($S=1$) final states, and serves to explain 
the near absence of singlet-state production seen in earlier 
measurements~\cite{PS185M,PS185_96}.
The QG approach, which includes only couplings to triplet-state 
quantum numbers, incorporates complete triplet dominance by construction 
while finding only a relatively small tensor interaction, even when fully 
accounting for the ISI and FSI.

In the measurement described here, 
the role of the tensor interaction in $\reac$ is probed through
measurements of the spin transfer observables, which describe correlations between
the initial and final state polarizations.  The depolarization $D_{nn}$ measures 
spin transfer from the target proton to the produced $\Lambda$, 
and is defined such that:
\begin{equation}
\langle \sigv_{\Lambda}\cdot\nh \rangle =
\begin{array}{c} P_{n} + D_{nn}\,\vpt\cdot\nh \\ 
\hline 1 + A_{n} \vpt\cdot\nh \end{array} .
\end{equation}
The polarization transfer $K_{nn}$ analogously measures the transfer of spin
from the target proton to the produced $\lbar$:
\begin{equation}
\langle \sigv_{\lbar}\cdot\nh \rangle =
\begin{array}{c} P_{n} + K_{nn}\,\vpt\cdot\nh \\
\hline 1 + A_{n} \vpt\cdot\nh \end{array} .
\end{equation}
In these expressions, $\langle \sigv\cdot\nh\rangle$ is twice the average
value of the spin component along the normal to the scattering plane, 
$\nh$. The direction of $\nh$ is defined in terms of the incident and
outgoing particle momenta, with $\nh$ in the direction of 
$\vp_{\bar{p}} \times \vp_{\lbar}$. 
Also, $\vpt$ is the target proton polarization, 
$P_{n}$ is the polarization of the $\Lambda$ and $\lbar$ in the $\nh$ direction
for production with no initial-state polarization, and 
$A_{n}$ is the left-right asymmetry of $\lbarl$ production with
a polarized target. 

Since the tensor interaction prefers spin-flip transitions between
the initial and final states, the MEX calculations 
predict a strongly negative $D_{nn}$ and $K_{nn}$.  In contrast,
the QG calculations include only a minor tensor component and consequently predict
less spin-flip and positive values for $D_{nn}$ and $K_{nn}$.  This difference in
the predictions of the models has been shown to be largely insensitive
to inclusion of the ISI and FSI~\cite{Haidenbauer,Alberg93}.

Previous PS185 results on spin observables have been limited to final-state
spins, which could be determined from event topology distributions since
the self-analyzing weak decay of the hyperon correlates the 
direction of the decay products with the hyperon spin.
The measurement of spin transfer observables 
requires the use of a polarized target or beam.  The experiment described in 
this Letter used a frozen spin target and 
represents the first measurement of such observables for exclusive
hyperon production from $\pbarp$ annihilation in the near-threshold 
region~\cite{PS185_3}.
It provides a stringent, new test for models which have successfully reproduced 
previous measurements of this reaction.

The detector system, which was essentially the same as that used for previous
PS185 measurements, has been described in several publications~\cite{PS185M}.  The 
products of the charged decay of the hyperons were tracked in 10
planes of multi-wire proportional chamber followed by 13 planes
of drift chamber.  The topology
of these four tracks, along with the well-known masses of the nucleons, 
hyperons, and pions, over-determines the kinematics of the event.  A fit of
the kinematics of the reaction to this topology provides a precise measurement
of the center-of-mass production and decay angles, as well as a clear method 
for distinguishing the signal events from background through the fit quality.  
There were no magnetic fields in these tracking chambers, leaving an ambiguity 
between the $\lbar$ and $\Lambda$. This ambiguity was 
resolved by using three additional drift planes to detect the horizontal 
deflection of each track in a vertical magnetic field contained in
a solenoid behind the tracking chambers.

The trigger system, which took advantage of the 
charged-neutral-charged signature of the 
event topology, was also similar to that used in 
previous PS185 measurements, although it required modification to accommodate the
frozen spin target.  The trigger was initiated by scintillators upstream
of the target, which detected the incident $\bar{p}$.  The trigger was
vetoed by scintillators downstream and to the sides of
the target, thus requiring neutral particles exiting the target, 
consistent with $\lbarl$ production.
The trigger was completed by coincidence with hits in a scintillator hodoscope
positioned downstream of the tracking chambers, which indicated the passage of
charged tracks through the active detector volume as expected for the 
charged decay of the hyperons.  

A transversely-polarized frozen-spin target~\cite{TargetNIM,TargetThesis} 
was used to provide access to the spin transfer observables. 
This target was a 6~mm diameter, 9~mm long cylinder of butanol submerged
in a liquid He bath, with the cylindrical axis aligned to the beam direction.
The cryostat,  which incorporated a superconducting solenoid to produce
the holding field, was a vertical cylinder with an outer diameter of only 42 mm.
This small cryostat size allowed the trigger veto scintillators 
to be positioned close to the production target.  
The close positioning of these scintillators was critical
to maintain trigger efficiency, as any hyperon which decayed
upstream of them caused the event to be vetoed.   The polarization of the 
target was determined using NMR measurements to fix the initial and final
points of the relaxation curve for each data-taking period~\cite{TargetNIM2}. 
The magnitude of target polarization averaged $62\%$ during data
production.

A detailed, GEANT-based Monte Carlo simulation~\cite{GEANT} 
of the detection and analysis
procedures was used to study the effects of imperfect geometric acceptance and 
reconstruction efficiency.  The detection efficiency correction was large, 
due primarily to the stringent trigger requirements. Over $80\%$ of the 
$\lbarl$ events that underwent doubly-charged decay were rejected
because of a decay upstream of the veto.
The effects of this trigger inefficiency were accurately reproduced by the
detector simulation, as confirmed by the extraction of proper lifetime 
distributions.  This simulation also included the effects of 
multiple Coulomb scattering of charged particles and hadronic interactions 
in the target and detector regions, which significantly affected the
event reconstruction efficiency.  

The target polarization sign was flipped
during data collection, so that approximately half of the total 
integrated luminosity was collected in each target polarization state
in order to control possible sources of systematic error.
A total of $1.2\times 10^{11}$ antiprotons were observed incident on 
the 9 mm thick target, producing 30,818 events which exhibited topology 
which could be cleanly fit by the kinematic hypothesis of
$\lbarl$ production from a free proton and subsequent doubly-charged decay.
These events were sorted into 16 bins of variable size 
over $\cos\theta\sscm$. The sizes of the bins were adjusted to approximately
match the statistics between them.  

The results presented in this Letter, for the observables
$\dnn$, $\knn$, and the singlet fraction $S_{F}$ as a 
function of the center-of-mass scattering angle $\cos\theta\sscm$, 
were extracted, simultaneously with other observables,
through the use of the spin scattering matrix formalism.  
Sensitivity to the parameters of the scattering matrix, which has 
recently been demonstrated~\cite{Paschke00}, depends on transverse
polarization of at least one initial-state particle as well as 
the self-analyzing property of the hyperon weak decay.  
The statistical error estimates for these results 
were determined by finding the limits of 
contours on the multi-dimensional log-likelihood function surface~\cite{Eadie}.
Due to correlations in parameter space, these error estimates are 
asymmetric for some observables in some bins of $\cos\theta\sscm$.
Details of the extraction of scattering matrix parameters and the 
uncertainty analysis, along with results for a larger set of observables, 
will be presented in a future publication.
For those observables which could also be extracted from previous
measurements performed without target polarization, good
agreement is found between the published results~\cite{PS185_96} 
and the corresponding observables found from this fit of the complete
set of parameters of the spin scattering matrix.

\begin{figure} \centering
\subfigure{
    \epsfig{file=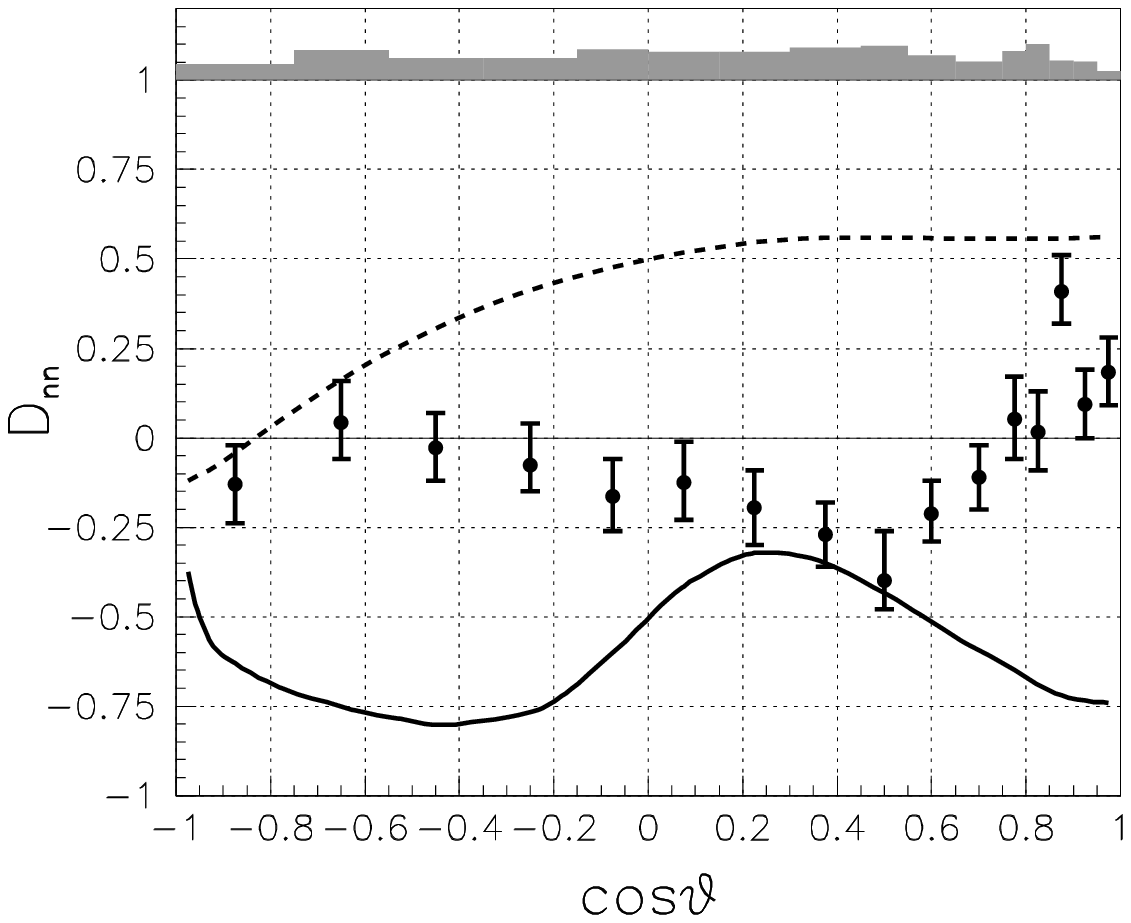,clip=,width=8.0cm}}
\subfigure{
    \epsfig{file=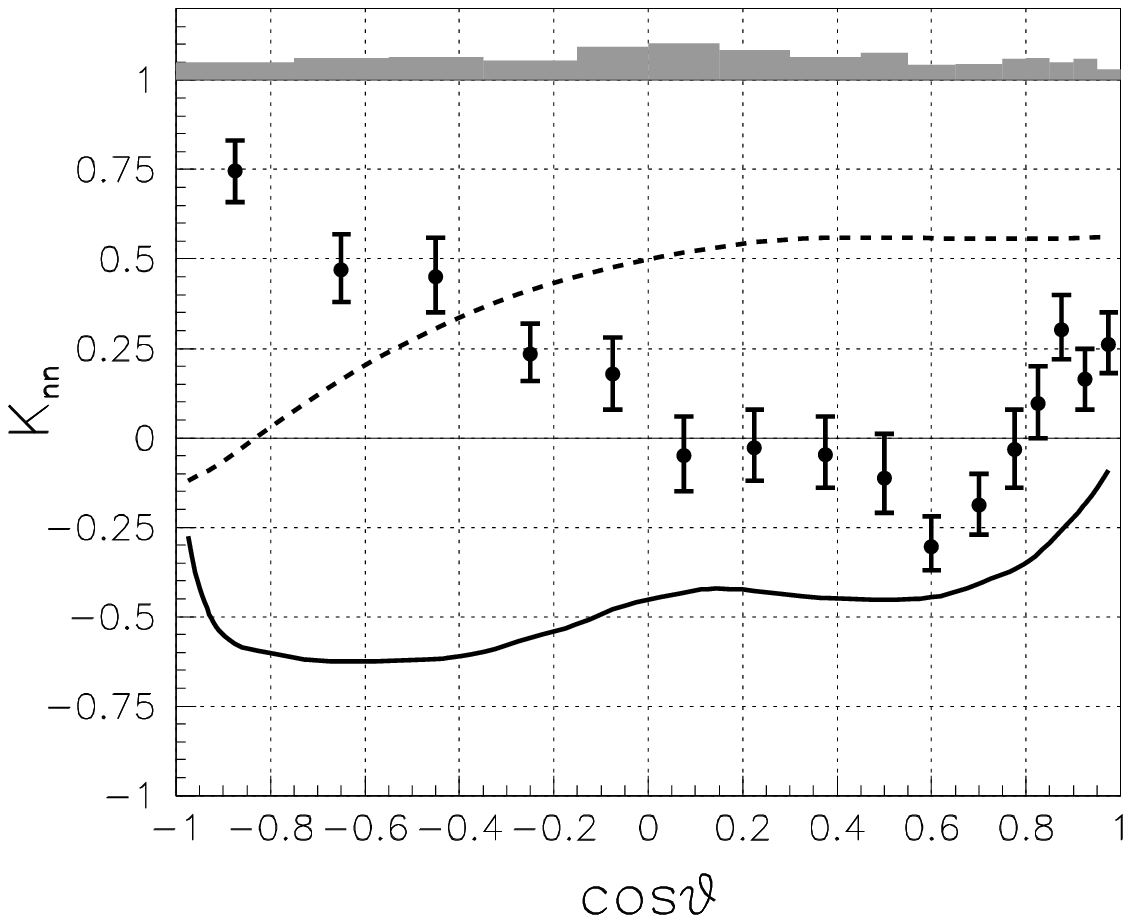,clip=,width=8.0cm}}
    \caption{Results for $D_{nn}$ and $\knn$
for $\reac$ at $1.637\pGeV$. The error bars indicate 1$\sigma$ statistical
uncertainty estimates, and are asymmetric for some data points.
The systematic uncertainty estimates are shown by the 
shaded boxes at the top of the figure.  
Predictions from the QG model~\protect\cite{Alberg93} (dotted) and the MEX 
model~\protect\cite{Haidenbauer} (solid) are superimposed.
\label{fig:sto}}  \end{figure}

Figure~\ref{fig:sto} shows results for the depolarization
$\dnn$ and polarization transfer $\knn$, along with predictions from full
QG~\cite{Alberg93} and MEX~\cite{Haidenbauer} calculations which include
ISI and FSI effects.  
These results clearly indicate a failure of 
both models to correctly describe the reaction dynamics 
in this kinematic region.

In the backward scattering region, $\dnn$ remains near zero, indicating 
that the target proton and $\Lambda$ spin are uncorrelated along the
scattering plane normal.  This contrasts with $\knn$, which 
grows from near zero at \mbox{$\cos\theta\sscm = 0$} to an average value
of \mbox{$\sim0.75$} for \mbox{$\cos\theta\sscm < -0.75$}, indicating a 
strong, positive
correlation between the target proton spin and the spin of the $\lbar$. 
The positive correlation between the initial-state baryon and the
final-state antibaryon in this range of \mbox{$\cos\theta\sscm$} 
was not predicted by either the MEX or the QG models.  

\begin{figure} \centering
    \epsfig{file=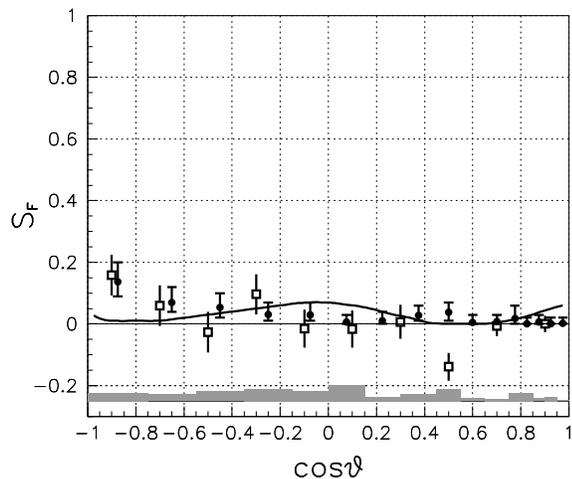,clip=,width=8.0cm}
    \caption{Current results for the singlet fraction $S_{F}$ 
for $\reac$ at $1.637\pGeV$ (solid circles) shown with asymmetric
statistical error bars, superimposed with results
from a previous measurement performed with an unpolarized target
at $1.642\pGeV$ (hollow squares)~\protect\cite{PS185_96} shown
with symmetric statistical error bars. 
The systematic uncertainty estimate
for the current measurement is shown by the shaded boxes at the 
bottom of the figure.  Predictions from
the MEX  model~\protect\cite{Haidenbauer} (solid) are superimposed.
The QG model predicts \mbox{$S_{F}=0$} uniformly~\protect\cite{Alberg93}.
 \label{fig:singf} }  \end{figure}

It is straightforward to demonstrate, using the spin scattering matrix formalism,
that $\dnn$ and $\knn$ are restricted to be equal in the case of
pure triplet-state production of the $\lbarl$ final state~\cite{Richard2}.
Therefore, the observed deviation between 
$\dnn$ and $\knn$ serves to emphasize the significance of
the small but non-zero singlet contribution in the final state.
The relative strength of the singlet and triplet components 
defines the Singlet Fraction observable, $S_{F}$, such that:
\begin{equation}
\langle \sigv_{\lbar}\cdot\sigv_{\Lambda} \rangle
= \begin{array}{c} 1 + 4 S_{F} \\ \hline
1 + A_{n} \vpt\cdot\nh \end{array} . \end{equation}
$S_{F}=1$ would signify pure singlet-state production while $S_{F}=0$ 
would imply pure triplet-state production.
This observable has been measured, by previous PS185 studies performed
with an unpolarized target, to be near zero in the near-threshold kinematic
region.
Measurements for $S_{F}$ have often been quoted as an average over center-of-mass 
production angle $\theta$, in which the strongly forward-peaked cross-section
weights the singlet-dominated forward-angle production more heavily.
Results for this observable from the current analysis as a function
of $\cos\theta\sscm$ are shown in Figure~\ref{fig:singf},
along with results from a previous measurement at similar kinematics.
There is good agreement between results from the two measurements, which
indicate that $S_{F}$ is very near zero for production in the forward
direction and is small, but non-zero, for back-angle production.  

This behavior is not well described by either the MEX or the QG model.
Without a small but
significantly non-zero singlet state component in the back-angle production,
no description of the production dynamics will be able to accommodate
the observed deviation between $\dnn$ and $\knn$.  In the case of the
QG model, only triplet-state transitions are allowed so the model, by
construction, describes a vanishing singlet fraction for all values of 
$\cos\theta\sscm$.  It is possible that the QG description could be 
improved with the inclusion of some singlet-state coupling, such as the
pseudoscalar $^{1}S_{0}$ transition, although previous studies have 
concluded that this transition was insignificant~\cite{Alberg91}.
While the MEX description is not so extreme with regard to the singlet contribution,
it also does not correctly describe the rising back-angle singlet fraction 
and so fails to predict the deviation between $\dnn$ and $\knn$.  In principle,
these models can accommodate an increased singlet contribution.  
Even with a somewhat reduced triplet-state strength, the MEX approach 
would tend to predict a dominant spin-flip between the initial and 
final states, which has been clearly excluded by the current measurements.

There has been some recent theoretical speculation about the 
connection between the near-threshold $\bar{Y}Y$ production
studies and the non-valence quark contributions to the 
nucleon spin~\cite{Alberg95,Ellis00}.  Mechanisms for describing
$\reac$ have been discussed which involve pre-existing polarized
strange quarks or polarized glue in the nucleon wavefunction. Although
no quantitative predictions have been made for $\dnn$ or $\knn$ 
with these models, the results presented here are not consistent with
the qualitative predictions from either of these alternative production 
mechanisms alone~\cite{Alberg95}.

In addition to the observables presented here, numerous other observables 
have been extracted, as has the full set of parameters of the spin scattering 
matrix~\cite{myThesis}, which will be shown in a future publication.
The breadth of these results present an opportunity for the further 
development of models describing the dynamics of the $\reac$ reaction.

\begin{acknowledgments}
The members of the PS185 collaboration thank the LEAR accelerator team.
We also gratefully 
acknowledge financial and material support from the German Bundesministerium 
f\"{u}r Bildung und Forschung, the Swedish Natural Science Research Council, 
the United States Department of Energy under contracts DE-FG02-87ER40315
and DE-FG03-94ER40821, and the United States National Science Foundation. 
\end{acknowledgments}


\end{document}